\documentclass[prl,twocolumn,eqsecnum,nofootinbib]{revtex4}
\usepackage{bm}
\usepackage{amssymb}
\usepackage{graphicx}
\usepackage{epstopdf}
\usepackage{mathrsfs}
\usepackage{amsmath}
\usepackage[hide]{errata}

\begin{document}
\title{Cauchy-horizon singularity inside perturbed Kerr black holes}
\author{Lior M.~Burko$^{1}$, Gaurav Khanna$^2$ and An{\i}l Zengino\v{g}lu$^3$}
\affiliation{$^1$ School of Science and Technology, Georgia Gwinnett College, Lawrenceville, Georgia 30043 \\
$^2$ Department of Physics, University of Massachusetts, Dartmouth, Massachusetts  02747\\
$^3$ Center for Scientific Computation and Mathematical Modeling, University of Maryland, College Park, MD 20742}
\date{June 4, 2015; Revised November 20, 2015}
\begin{abstract} 
The Cauchy horizon inside a perturbed Kerr black hole develops an instability that transforms it into a curvature singularity. We solve for the linearized Weyl scalars $\psi_0$ and $\psi_4$ and for the curvature scalar  $R_{\alpha\beta\gamma\delta}R^{\alpha\beta\gamma\delta}$ along outgoing null rays approaching the Cauchy horizon in the interior of perturbed Kerr black holes using the Teukolsky equation, and compare our results with those found in perturbation analysis. Our results corroborate the previous perturbation analysis result that at its early parts the Cauchy horizon evolves into a 
deformationally-weak, null, scalar-curvature singularity. 
We find excellent agreement
for $\psi_0(u={\rm const},v)$, where $u,v$ are advanced and retarded times, respectively. We do find, however, that  the exponential growth rate of $R_{\alpha\beta\gamma\delta}R^{\alpha\beta\gamma\delta}(u={\rm const},v)$ approaching the singularity is dramatically slower than that found in perturbation analysis, and that the angular frequency is in excellent agreement.

\end{abstract}
\maketitle

A singularity evolves inside every black hole (BH), as guaranteed under very plausible conditions by the Hawking-Penrose singularity theorems \cite{hawking-penrose}. Relatively little is known about the nature of BH singularities, including their physical and geometrical properties, the possibility of extension of the spacetime manifold beyond them, or the role played by quantum gravity. 

Based on properties of spherical toy models and of cosmological singularities in addition to perturbative and non-perturbative analysis of rotating BHs, it is believed that three types of singularities could exist, and possibly co-exist, inside realistic BHs: first, a BKL-type singularity \cite{bkl}, a spacelike,  anisotropic, homogeneous and chaotic singularity, in which Kasner epochs alternate independently along different timelike approaches to the singularity; second, the Poisson-Israel mass-inflation singularity \cite{poisson-israel}, a null and deformationally-weak singularity that evolves along the generators of the BH's Cauchy horizon, or ingoing inner horizon, because of the capture of future perturbations; and third, the Marolf-Ori singularity \cite{marolf-ori}, a null shock-wave singularity that evolves along the generators of the outgoing inner horizon because of the capture of past perturbations.

Here we are mostly concerned with the mass-inflation singularity inside rotating BHs, and specifically with the properties of the singularity in its early parts. 
The mass-inflation singularity was simulated in fully nonlinear simulations for a spherical charged scalar-field toy model in Ref.~\cite{brady-smith}, where the properties of its general features were confirmed. However, several key details of the features of the mass-inflation singularity that were predicted by perturbative analysis \cite{ori-91} appeared to be inconsistent with the nonlinear results of \cite{brady-smith}. This discrepancy was resolved by the careful simulations of Ref.~\cite{burko-97}, which showed that in the early parts of the mass-inflation singularity perturbation analysis is highly reliable and accurate. The picture of the singularity structure in spherical, charged, scalar-field toy models is that of a spherical, null, weak singularity, which is accurately described by perturbation analysis at early times, and that propagates at the speed of light while its null generators  contract, until eventually it contracts to zero volume when the singularity becomes spacelike. The spacelike singularity found in spherical charged toy models is strong,  scalar-curvature, and monotonic \cite{burko-99}, and is very different from the BKL singularity. The occurrence of the mass-inflation singularity in spherical symmetry was shown rigorously in \cite{dafermos}. 

Realistic black holes are neither spherical nor charged. The evidence for the singularity structure inside rotating black holes is made mostly of linear and nonlinear perturbation analysis \cite{ori-92,ori-99,ori-99a}, in addition to non-perturbative studies \cite{brady-chambers,ori-flanagan,ori-1998}. The early parts of the mass-inflation singularity inside asymptotically-flat, vacuum, rotating BHs is the main topic of the present paper. We evolve a linearized gravitational field over a Kerr spacetime using the Teukolsky equation, and find the behavior of the (azimuthal modes of the) linearized Weyl scalars $\psi_0$ and $\psi_4$ in the Kinnersley tetrad, and of the linearized curvature scalar $K:=R_{\alpha\beta\gamma\delta}R^{\alpha\beta\gamma\delta}$ approaching the mass-inflation singularity along an outgoing null direction. Since our code is a linear one, this paper cannot address inherently nonlinear phenomena such as the contraction of the null generators of the Cauchy horizon, the formation of a possible spacelike singularity, or the properties thereof. We can find, however, much about the early parts of the mass-inflation singularity. As detailed below, we first check, verify, or critique results found by perturbative \cite{ori-99} and non-perturbative \cite{brady-chambers} analyses, and then go beyond the latter and find additional phenomenology. This is the first numerical simulation for the dynamical  evolution of the gravitational field inside a perturbed spinning BH. 
Addressing the nonlinear questions requires nonlinear simulations of the full Einstein equations. With the coming of age of numerical relativity it is hoped that a new era of numerical geometrodynamics will commence. 

In our numerical experiments we find the Teukolsky equation to be numerically unstable when expressed in the common Boyer-Lindquist  or ingoing Kerr coordinates. Indeed, already in Schwarzschild spacetime the Teukolsky equation for $\Psi_0$, for spin $s=2$, in numerically unstable. (In Minkowski spacetime, it is the equation for $\Psi_4$, for spin $s=-2$, which is unstable.) This numerical instability comes about because of the presence of terms in the Teukolsky equation that effectively act as anti-damping terms. This instability presents us with a challenge when exploring the singularity of a Kerr BH.

\begin{figure}
\includegraphics[width=7.5cm]{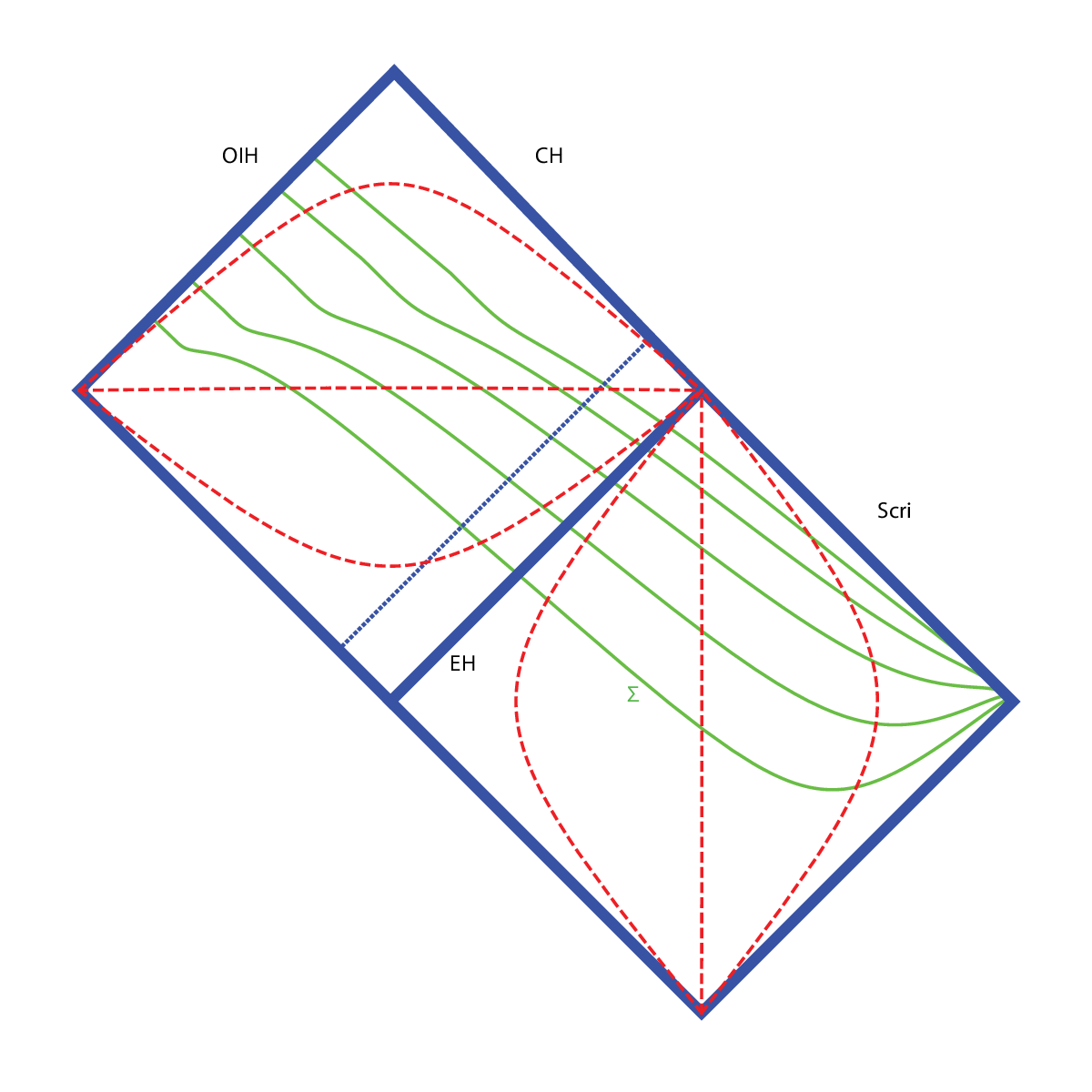}
\caption{The Penrose diagram for the domain of integration and the coordinates. The solid (dashed) thin curves are hypersurfaces $\tau={\rm const}$ ($\rho={\rm const}$). Notice that the $\tau={\rm const}$ surfaces intersect  on the diagram with $\mathscr{I}^+$, and penetrate through the event horizon (EH) and the outgoing inner horizon (OIH), which is the inner boundary for our computational domain. The earliest of these surfaces is our partial Cauchy surface $\Sigma$. The dotted thin line is a  constant advanced-time ray ($u={\rm const}$) that intersects with the Cauchy horizon (CH).}
\label{Penrose}
\end{figure}

Denoting the ingoing Kerr coordinates (${\tilde t},r,\theta,{\tilde\varphi}$) we define the dimensionless compactified hyperboloidal coordinates $(\tau,\rho,\theta,{\tilde\varphi})$ by $\tau := {\tilde{t}}' - {r'}^2/({r'}+S) + 4 \ln [S/({r'}+S)]$ and $\rho :={r'}/[1+{r'}/S]$, where ${\tilde{t}}':=\tilde{t}/M$ and ${r'}:=r/M$. The free parameter $S$ controls the numerical domain and the foliation, and gives us some freedom in the number of grid points and the size of the time steps  \cite{harms-2014}.  In these coordinates the horizons are located at 
$\rho_\pm = \frac{ a^2 S +  S^2M^2 \pm S^2 M\sqrt{M^2 - a^2} } { a^2 + 2  S M^2 + S^2 M^2}\,,$ and the PDE becomes mixed type hyperbolic-elliptic. Notice that $\rho\in [0,S)$ maps   the domain $r\in [0,\infty )$ one-to-one. Figure \ref{Penrose} shows the coordinates on a Penrose diagram of a Kerr BH. 

\begin{figure}
\includegraphics[width=7.5cm]{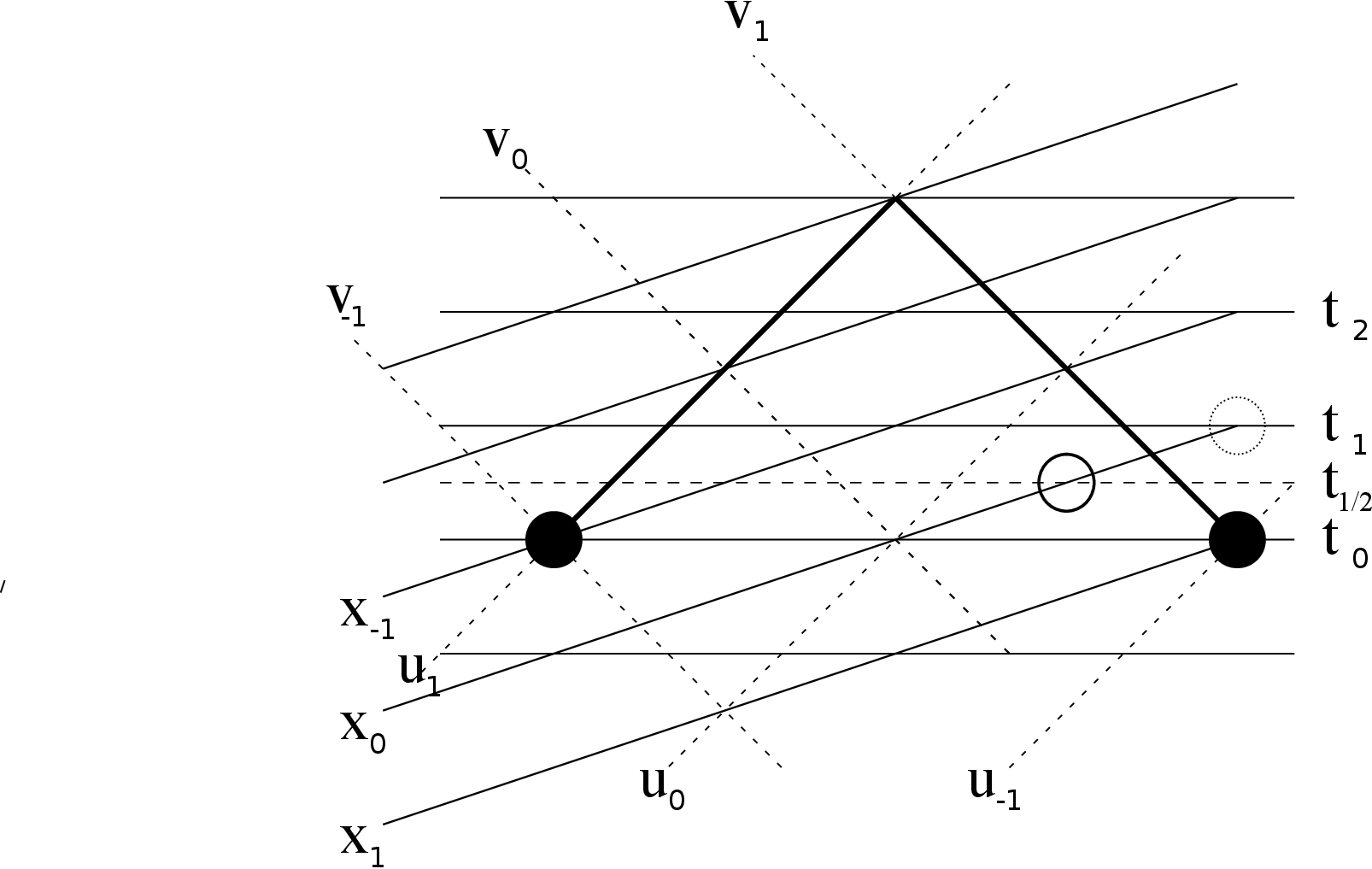}
\caption{Sketch of the numerical scheme for Cauchy data with two timelike coordinate ($t,x$). Cauchy data are specified for $t_0$ at $x_{-1}$ and at $x_1$. With one spacelike $x$ coordinate one normally finds the fields at $(t_1, x_0)$. Here, that event is outside the domain of influence of the initial data. One may, however, find the fields at, say, $(t_{1/2},x_0)$, which is inside the domain of influence.  Shown also are retarded and advanced times.}
\label{coordinates}
\end{figure}

\begin{figure}
\includegraphics[width=7.5cm]{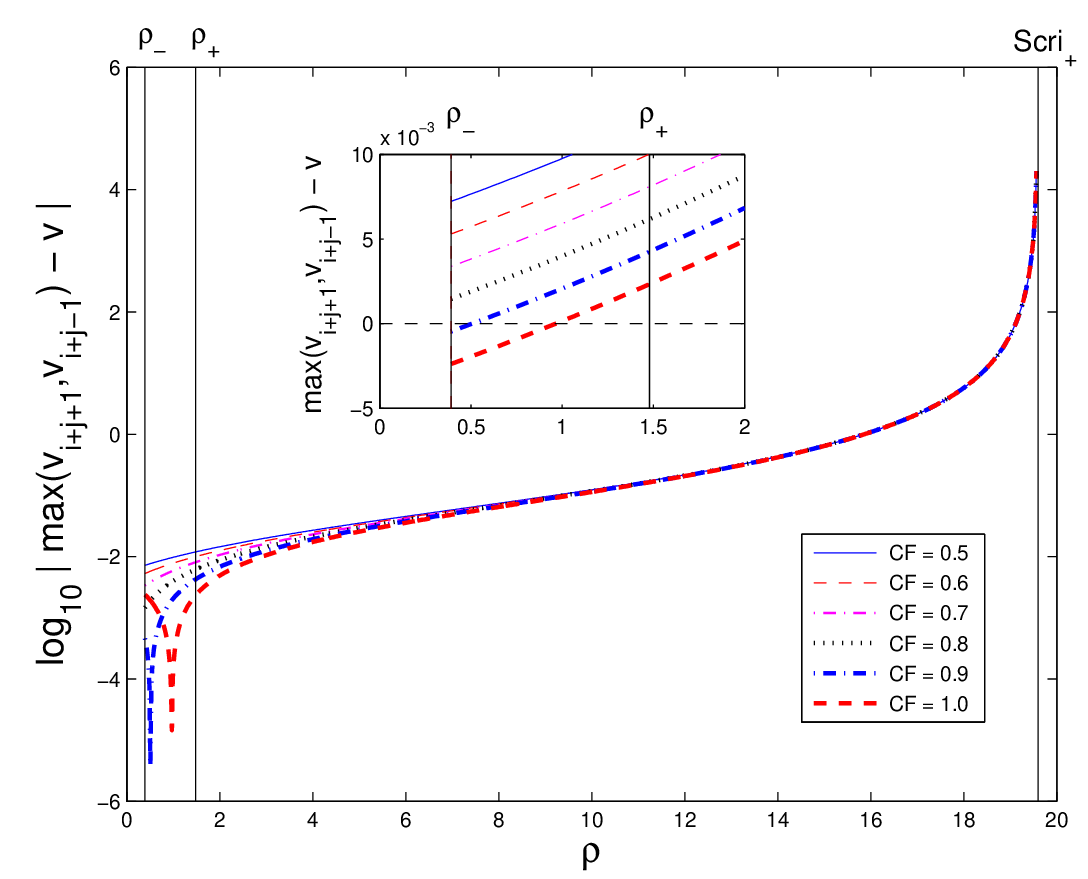}
\caption{ The left hand-side of Inequality (\ref{courant}) as a function of $\rho$ for various values of the Courant factor, for a Kerr BH with $a/M=0.8$. The horizons are at $\rho_+=1.479$ and $\rho_-=0.392$ ($S=19.6$). When the Courant factor is smaller than $\sim 0.8$ the CFL condition is globally satisfied.  The condition on advanced time is satisfied globally for all Courant factors $\le 1$.}
\label{CFL}
\end{figure}

Inside the BH our coordinates have the unusual signature $(-,-,+,+)$, and our computation method is solving an initial value problem using Cauchy data. While the usual situation is that the metric signature is $(-,+,+,+)$ (``one time dimension"), one may still set up the evolution scheme with our unusual signature. 
Figure \ref{coordinates} shows how the Courant-Friedrichs-Lewy (CFL) condition can be satisfied even when all equal-coordinate surfaces are spacelike (in a 1+1D toy model). Our numerical scheme satisfies the CFL condition in practice. Part of the CFL condition is that if $v_{i+j-1}$ and $v_{i+j+1}$ are the retarded times of the two data points on which Cauchy data are specified, and $v$ is the retarded time at which we want to find the field, than the CFL condition requires that 
\begin{equation}{\rm max}(v_{i+j-1},v_{i+j+1})-v>0
\, . \tag{1}
\label{courant}
\end{equation} 
(A similar relation hold also for advanced time $u$.) Figure \ref{CFL} shows the behavior of the CFL condition as a function of $\rho$. Clearly, we can choose the Courant factor small enough so that the CFL condition is globally satisfied in the domain of integration. The CFL condition is of course only a necessary condition. In addition to showing that the CFL condition is satisfied, we have also tested numerically the stability and convergence of the solution. 

There are several advantages offered by the coordinates 
$(\rho,\tau,\theta,{\tilde\varphi})$: First, they provide a clean solution to the so-called 
``outer boundary problem." One major challenge in the numerical solution 
of a PDE arises from the fact that the computational domain needs to be 
finite. Well-posed boundary conditions along with transparent boundary data are required at the boundary, which are difficult to construct and implement numerically. Standard methods for boundary treatment lead to spurious ÒreflectionsÓ from the outer boundary and contaminate the solution. The standard 
approach currently in use by many communities is to place the outer boundary 
far enough that these reflected spurious  
waves are unable to interfere with the physical solution. This approach makes the computational domain unnecessarily large and the computation very expensive. With the approach of compactification of the computational domain, one is able to 
extend the domain to infinity, thus offering a clean solution to this 
problem. Second, because the compactified computational domains have a 
rather modest extent (in our simulations we set $S\sim 20$)  one has the ability 
to perform computations with grid-resolutions that are enormously high at 
very low cost. For instance, in this work we use uniform grid-spacings (in $\rho$) on the scale
of $1/2,000$. This dense computational grid allows us to obtain results which are accurate to $1\%$. 
(In addition, we take full advantage of GPU-accelerated computing to perform our computations efficiently.)  
Third, the compactification allows us to extract signals at  (future null) infinity ($\mathscr{I}^+$) directly, because it is part of the computational domain. And finally, these coordinates also resolve the aforementioned instability problem. See Ref.~\cite{ZengKha:2011} for more detail. 

Notably, our coordinates compactify $\mathscr{I}^+$, thereby allow us to compactify the domain of computation without introducing reflection of the waves back into the interior region. This way, we  avoid one of the main difficulties associated with the compactification of spatial infinity. In fact, our hypersurfaces are different from the Boyer--Lindquist hypersurfaces in such a way that approaching $\mathscr{I}^+$ the spatial wavelength of outgoing radiation is unbounded in our foliation.
(``$\tau={\rm const}$ hypersurfaces are asymptotically wave fronts"),
Therefore we can resolve the field on the compactified grid and do not encounter spurious reflections due to compactification near the outer grid boundary \cite{anil-2011}. 

\begin{figure}
\includegraphics[width=7.5cm]{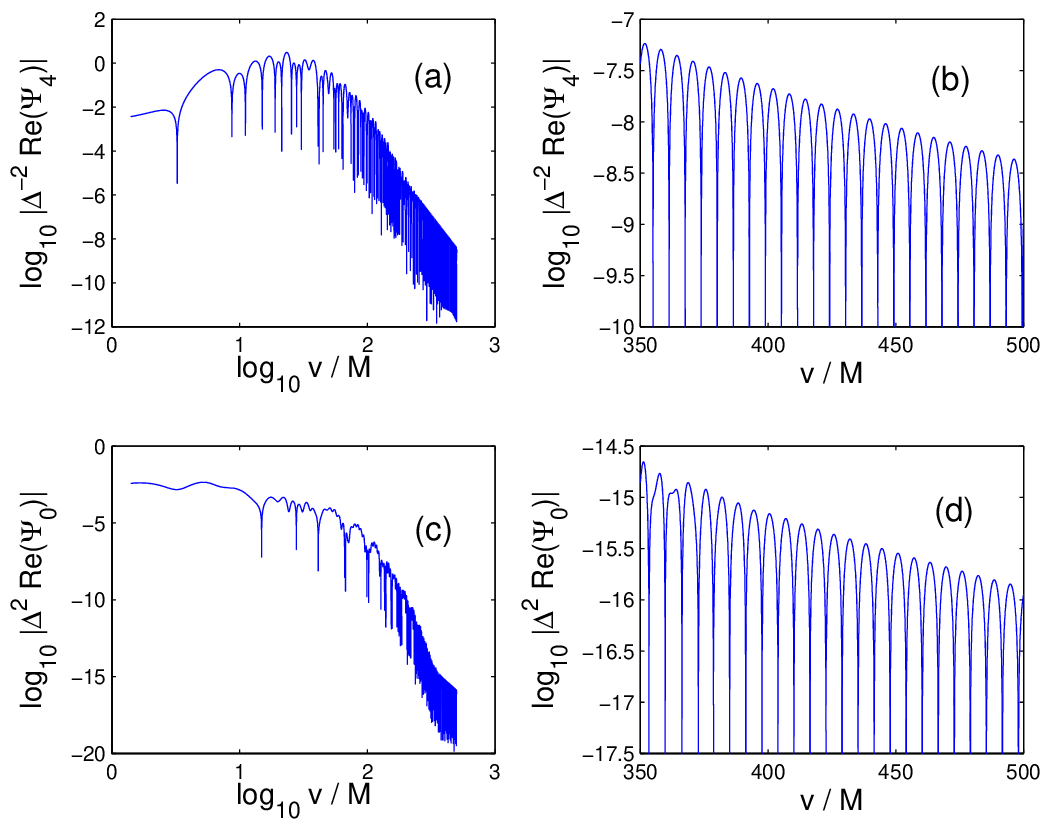}
\caption{The Teukolsky fields $\Delta^{-2}\Psi_4$ (panels (a) and (b)) and $\Delta^{2}\Psi_0$ (panels (c) and (d)) along the event horizon on the equatorial plane  as functions of retarded time for $a/M=0.8$ and $m=2$ in horizon regularized coordinates.}
\label{eh}
\end{figure}

A 2+1D code for the $m$ modes of $\Psi_0$ and $\Psi_4$ based on the ($\tau,\rho,\theta,{\tilde\varphi}$) coordinates (and which solves for the fields as functions of $(\tau,\rho,\theta)$),    
is numerically stable and convergent for both fields in the entire domain of interest \cite{RT}, and reproduces results in agreement with known or expected ones, such as the behavior of the fields along the event horizon, as demonstrated in Fig.~\ref{eh}. These are the first numerical results for both the decay rate and the oscillation frequency for the fields along the event horizon for non-axisymmetric gravitational perturbations.

We transform the azimuthal coordinate 
$\,d\tilde\varphi\to\,d\tilde\varphi_{\pm}:=\,d\varphi-\Omega_{\pm}\,dt$, where $t,\varphi$ are the Boyer-Lindquist coordinates. The horizon frequency $\Omega_{\pm}=a/(2Mr_{\pm})$. 
The horizon-regularized coordinates (${\tilde t},r,\theta,{\tilde\varphi}_{\pm}$) are appropriate coordinates for the description of fields interacting with observers, and also allow us to directly compare our results with those of \cite{ori-99,ori-99a,brady-chambers}.

We note that while a scalar $Z$ (such as $\psi_0$, $\psi_4$, or $K$) is invariant under coordinate transformations, its $m$ modes are not. That is, we decompose $Z$ in our code as 
$Z=\sum_{m=0}^{\infty}Z^{(m)}\,e^{im{\tilde\varphi}}\, ,$
but we wish to present our results for the modes 
$Z=\sum_{m=0}^{\infty}Z^{(m)}_{\pm}\,e^{im{\tilde\varphi}_{\pm}} \, .$
The two decompositions are related by
$\,d{\tilde\varphi}=\,d{\tilde\varphi}_{\pm}+\Omega_{\pm}\,dv$, such that 
$Z^{(m)}_{\pm}=Z^{(m)}  e^{im\Omega_{\pm}v}\, ,$ where $v$ is retarded time. 


Perturbation analysis predicts the asymptotic behavior of $\psi_0(u={\rm const},v)$ approaching the Cauchy horizon. Specifically,  Refs.~\cite{ori-99,ori-99a} found that as $v/M\to\infty$ 
\begin{equation}
\psi_0(u={\rm const},v)\sim \Delta^{-2}\,e^{i\omega_{\rm p} v}\,v^{-\alpha_{\rm p}}\left[1+O(v^{-1})\right]\, , \tag{2}
\end{equation}
where 
$\omega_{\rm p}=m\Omega_-$,  and where for the fastest-growing gravitational mode $\alpha_{\rm p}=7$. Here, $\Delta:=(r-r_+)(r-r_-)$, and $\log(\Delta/r^2)\propto -v/M$ as $v/M\to\infty$. (Ref.~\cite{brady-chambers} does not calculate the frequency.) 
We note that \cite{ori-99} does not make a prediction for $\psi_4(u={\rm const},v)$ (although it does make a prediction for $\psi_4(u,v={\rm const})$.)

We next present the behavior of fields approaching the Cauchy horizon along an outgoing null direction. Figure \ref{psi04} shows the behavior of the Weyl scalars $\psi_0$ and $\psi_4$ as functions of retarded time. Note that the numerical solution solves for the fields starting at magnitude $O(1)$ and ending at $O(10^{\sim 300})$ close to the Cauchy horizon. In Table \ref{table:psi0} we parametrize our solution for $\psi_0$. We find it to agree well with that of perturbation analysis in both the power-law decay rate and its frequency. Note that our numerical solution has a global phase difference compared with the solution of \cite{ori-99}, which results from a shifted origin of retarded time, but no dephasing. 

\begin{figure}
\includegraphics[width=7.5cm]{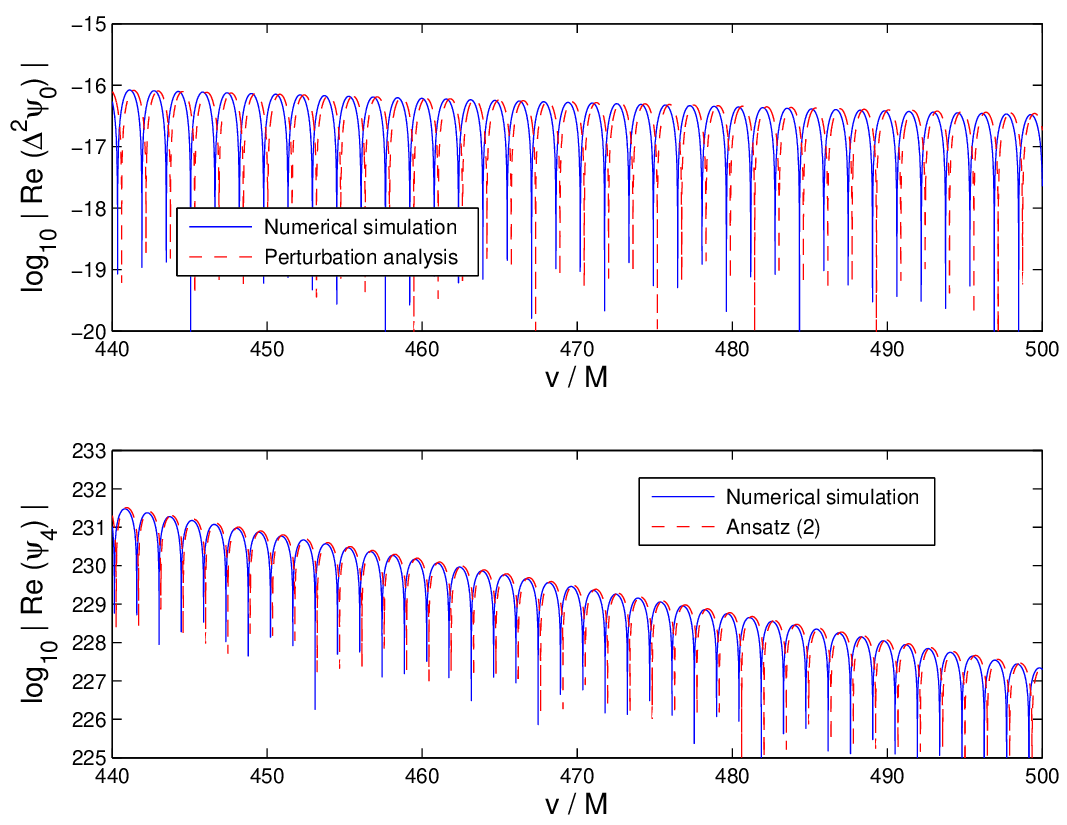}
\caption{The Weyl scalars $\Delta^2\psi_0$  in regularized coordinates (upper panel) and $\psi_4$ in ingoing Kerr coordinates (lower panel) as functions of $v/M$ along an outgoing null ray that intersects with the early part of the Cauchy horizon (as $v/M\to\infty$). For $\Delta^2\psi_0$ ($\psi_4$) we present in addition to our numerical results in a solid curve also the prediction of perturbation analysis (Ansatz (\ref{ansatz})) in a dashed curve.  
Here, $a/M=0.8$ and $m=2$ and the fields are on the equatorial plane.}
\label{psi04}
\end{figure}

We also solved for $\psi_4(u={\rm const},v)$. Here we have no predictions from perturbation analysis with which to confront our numerical results. Instead, we postulate an Ansatz, and find the parameters that best fit it. Specifically, we postulate that in ingoing Kerr coordinates 
\begin{equation}\label{ansatz}
\psi_4(u={\rm const},v)\sim e^{-i\omega v}\, e^{-v/mW}\, . \tag{3}
\end{equation}
The lower panel of Fig.~\ref{psi04} and Table \ref{table:psi4} show our solution for $\psi_4$. Our data suggest  that the angular frequency $\omega=m\Omega_-$, and that the phenomenological free parameter $W$ increases rapidly with $a/M$. Therefore, in horizon regularized coordinates $\omega=0$, and $\psi_4$ does not oscillate approaching the Cauchy horizon. 

\begin{table}
\caption{Parameters for the numerically simulated $\psi_0$ approaching the Cauchy horizon in regularized coordinates: for $a/M=0.8$ for various values of $m$ (left side), and for $m=2$ for various values of $a/M$ (right side). The parameters $\alpha$ and $\omega$ are found from the numerical simulations. The parameters $\alpha_{\rm p}$ and $\omega_{\rm p}$ are the predictions of perturbation analysis. Data, on the equatorial plane, are presented in horizon regularized coordinates.
}
\centering
\begin{tabular}{ || c | c | c | c | c || }
  \hline                       
  $m$   & $\alpha$ & $\omega M$ & $\alpha_{\rm p}$ & $\omega_{\rm p} M$  \\ \hline 
  1 & 7.19 & 1.00 &  7 & 1.000 \\  \hline 
    2 & 7.22 & 2.00 & 7 & 2.000 \\  \hline 
  3 & 8.83 & 3.00&  9 & 3.000  \\
  \hline  
\end{tabular}
\hspace{0.3cm}
\begin{tabular}{ || c | c | c | c | c || }
  \hline                       
  $a/M$   & $\alpha$ & $\omega M$ & $\alpha_{\rm p}$ & $\omega_{\rm p} M$  \\ \hline 
  0.800 & 7.22 & 2.00 &  7 & 2.000 \\  \hline 
    0.866 & 7.28 & 1.73 & 7 & 1.732 \\  \hline 
  0.917 & 7.26 & 1.53&  7 & 1.528  \\ \hline  
\end{tabular}
\label{table:psi0}
\end{table}

\begin{table}
\caption{Parameters for the numerically simulated $\psi_4$ approaching the Cauchy horizon: for $a/M=0.8$ for various values of $m$ (left side), and for $m=2$ for various values of $a/M$ (right side). The parameters $W$ and $\omega$ are found from the numerical simulations. The relative difference between $\omega M$ and $m\Omega_-M$ is denoted by $\delta$. 
Data, on the equatorial plane, are presented in ingoing Kerr coordinates.}
\centering
\begin{tabular}{ || c | c | c | c || }
  \hline                       
  $m$   & $W/M$ & $\omega M$ & $m\Omega_-M$  \\ \hline 
  1 & 3.68 &   1.15 & 1.000 \\  \hline 
    2 & 3.09 &  2.18 & 2.000 \\  \hline 
  3 & 3.64 &   3.17 & 3.000  \\
  \hline  
\end{tabular}
\begin{tabular}{ || c | c | c | c  | c || }
  \hline                       
  $a/M$   & $W/M$ & $\omega M$ & $m\Omega_-M$  & $\delta$ \\ \hline 
  0.800 & 3.09 & 2.18 &   2.000  & 0.083 \\  \hline 
    0.866 & 6.56 & 1.85 &  1.732 & 0.063 \\  \hline 
  0.917 & 16.5 & 1.60 &   1.528  & 0.047 \\ \hline  
   0.954 & 71.3 & 1.40 &   1.363  & 0.024 \\ \hline
    0.980 & $9\times 10^2$ & 1.25 &   1.225 & 0.017 \\ \hline
\end{tabular}
\label{table:psi4}
\end{table}



From the Weyl scalars $\psi_0$ and $\psi_4$ are next find the linearized curvature scalar  
$$K=8\left(\psi_0\psi_4+3\psi_2^2-4\psi_1\psi_3\right)+{\rm c.c}\, .$$
As pointed out in \cite{ori-99,brady-chambers}, $K$ is dominated approaching the Cauchy horizon by  $K\sim 8\psi_0\psi_4+{\rm c.c}$.

In Figure \ref{kretch} we present $K$ as a function of retarded time. Our previous results suggest to us that as $v/M\to\infty$, the fastest growing mode of
\begin{equation}
K(u={\rm const},v)\sim 
\Delta^{-2}\, e^{im\Omega_-v}\, e^{-v/mW}\,v^{-7}\, . \tag{4}
\end{equation}
The analysis of Ref.~\cite{ori-99} finds, however, that 
$$K_{\rm p}(u={\rm const},v)\sim 
\Delta^{-2}\, e^{im\Omega_-v}\,v^{-7}\, .$$
(See also \cite{brady-chambers}.) 
We find that the growth rates of curvature  are dramatically different: The exponential growth rate for curvature approaching the Cauchy horizon that we find is slower than that found in \cite{ori-99}. Therefore, the difference between the two expressions for the curvature grows exponentially with retarded time. 

We conjecture that the reason for this discrepancy between the perturbative results of Ref.~\cite{ori-99} and our results is that \cite{ori-99} finds $\psi_0(u={\rm const},v)$ and $\psi_4(u,v={\rm const})$, and uses these in order to calculate $K_{\rm p}(u={\rm const},v)$. This approach amounts to tacitly assuming that $\psi_4(u={\rm const},v)\sim{\rm const}$ asymptotically. As we show above in Fig.~\ref{psi04} and in Table \ref{table:psi4}, this tacit assumption is unrealized in our numerical simulations.

\begin{figure}
\includegraphics[width=7.5cm]{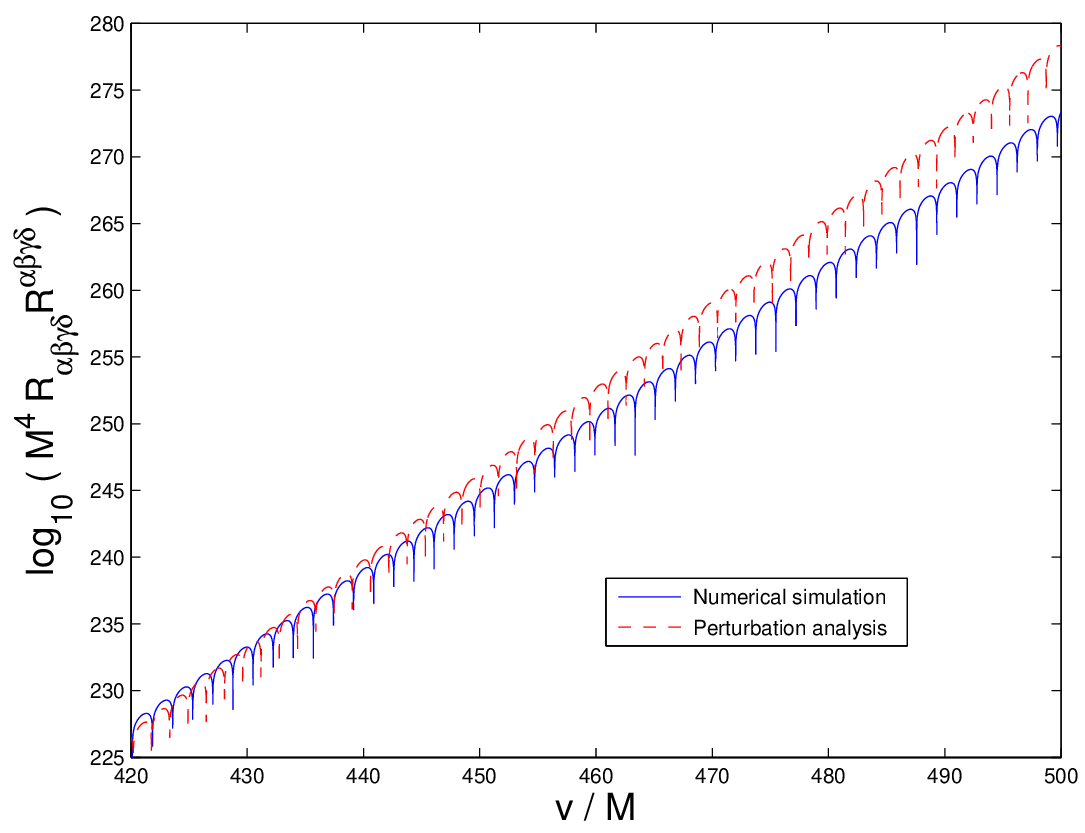}
\caption{The curvature scalar $R_{\alpha\beta\gamma\delta}R^{\alpha\beta\gamma\delta}$ as a function of $v/M$ along an outgoing null ray on the equatorial plane approaching the Cauchy horizon. Here, $a/M=0.8$ and $m=2$, and data are on the equatorial plane. The numerical simulation is shown with a solid curve, and the prediction of \cite{ori-99} with a dashed curve.}
\label{kretch}
\end{figure}

We note that this disagreement is based on the fit parameters for the Ansatz (\ref{ansatz}). While this Ansatz for the fit functions fits the data very well, the data represent a rather short interval of retarded time. Consider, say, the hypothetical alternative Ansatz $\psi_4(u={\rm const},v)\sim e^{-i\omega v}\, (e^{-v/mW}+A)$. If the parameter $A\ll 1$ it is hard to find numerically a non-zero value for $A$ unless the interval in retarded time is very long. In our numerical simulations it is difficult to significantly increase the retarded-time interval: the function we solve for, $\Delta^{-2}\Psi_4$, grows exponentially with retarded time, and floating-point arithmetics limits the interval. 

Consequently, our data represent evolutions to $v/M=500$. In the context of external  perturbations this length of evolution typically allows for only a crude estimate of power law indices, even when combined with an evaluation of the local power index \cite{burko-ori-97} and its extrapolation to infinite time. We believe this is the case also here, contributing to the disagreements reflected in Tables \ref{table:psi0} and \ref{table:psi4}. Specifically, our results do no depend on an Ansatz regarding the asymptotic form of the fields (such as assuming a Price law behavior at finite times along the EH), such that subdominant terms may still contribute. 

Even if the aforementioned retarded-time problem can be resolved, we face the fact that there is little  sense in continuing our evolution beyond the level that curvature become Planckian. While general relativistic nonlinear effects may be ignored at the early parts of the Cauchy horizon, we do not yet know what any quantum gravity effects may be. Our results may have important implications for the latter.

It will be interesting to use a similar code in order to study the behavior of fields along the Cauchy horizon, and also along and across the outgoing inner horizon. Further details of this work will appear elsewhere.

\section*{Acknowledgements} 
The authors are indebted to Richard Price for discussions. 
G.K.~acknowledges research support from NSF Grants No.~PHY--1303724 and No.~PHY--1414440, and from the U.S.~Air Force agreement No.~10--RI--CRADA--09.

\newpage
\onecolumngrid
\appendix

\noindent {\large{\bf Erratum: Cauchy-horizon singularity inside perturbed Kerr black holes [Phys.~Rev.~D {\bf 93}, 041501(R) (2016)]}}
\newline

\vspace{0.2 cm}

\centerline{Lior M.~Burko, Gaurav Khanna, and An{\i}l Zengino\v{g}lu}
\vspace{0.15 cm}

\centerline{(December 1, 2017)}

\vspace{0.2 cm}

In this Erratum we correct the behavior of the Weyl scalar $\psi_4$ and of the linearized curvature scalar $K:=R_{\alpha\beta\gamma\delta}R^{\alpha\beta\gamma\delta}$ approaching the mass-inflation singularity along an outgoing null direction. In the original paper we used a second-order code, that allowed us to find very accurately the fields in the exterior and on the event horizon. However, it has some disadvantages approaching the Cauchy horizon (CH). Specifically, our code solves for $\Delta^{-2}\,\psi_4$, which diverges exponentially with advanced time approaching the CH (more specifically, diverges like $e^{2v/M}$). Moreover, in the code's coordinates this field oscillates approaching the CH. To increase accuracy, we developed a new fifth-order WENO finite-difference scheme \cite{WENO} with third-order Shu-Osher explicit time-stepping \cite{gotlieb-shu-tadmor,gotlieb-book}, which we also used in \cite{burko-khanna}. More importantly, we improved the treatment of the inner boundary of the computational domain, which is the outgoing leg of the inner horizon (OIH) (see Fig.~1). Specifically, the fields are now actually ``evolved'' on the inner boundary as opposed to computed using the boundary conditions in conjunction with data from the ``bulk''. In the original paper we used the latter approach.  As we expect sharp physical features to be present at the boundary  \cite{marolf-ori}, the approach used here allows for more accurate numerical solutions. In practice, the source-side of the Teukolsky equation is computed at the boundary and the field values are updated at every time-step. Notice that computing the source-side involves computing derivatives at the boundary, and that is done using a high-order, one-sided, finite-difference stencil. (See \cite{burko-khanna} for more detail.)

This new computational approach allowed us to resolve the apparent disagreement between our results for $\psi_4$ and $K$ and the predictions of perturbation theory \cite{ori-99}: we now show agreement in both frequency and magnitude. 

Specifically, in the original paper we introduced a fit parameter $W$ that controls the asymptotic deviation of $\psi_4$ from a limiting constant magnitude. An  asymptotically constant-magnitude $\psi_4$ would then be obtained in the limit $W\to\infty$ as advanced time $v\to\infty$. With our improved code we find that $W$ would take very high values. (E.g., for $a/M=0.8$ and $m=1$, we find that $W\sim 1,400M$.) We therefore conclude that we no longer have a reason to introduce $W$, and that in practice $W\to\infty$. To illustrate this result, we present in Fig.~\ref{constancy} the asymptotic behavior of $\psi_4$ in ingoing Kerr coordinates for a number of $a/M$ and $m$ values. 

\setcounter{figure}{6}

\begin{figure}[h]
\includegraphics[width=7.5cm]{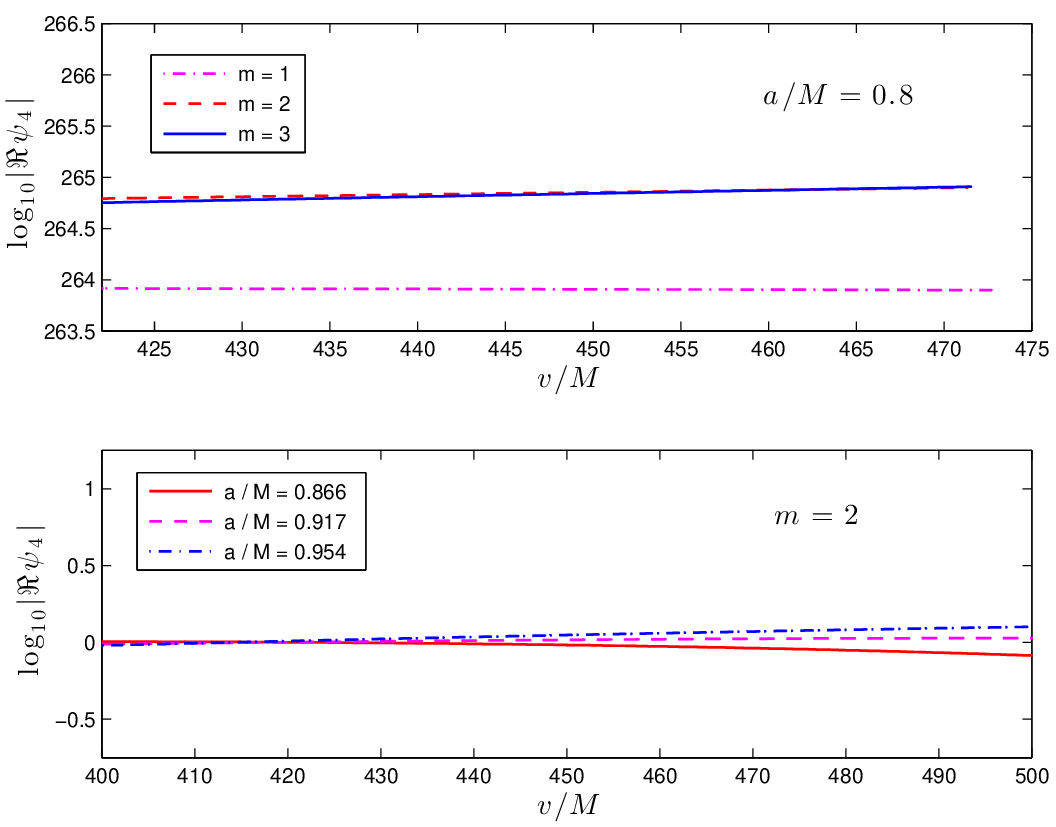}
\caption{The magnitude of (the real part of) $\psi_4$ in ingoing Kerr coordinates as a function of advanced time along an outgoing null ray. Top panel: The cases $m=1$ (dash-dotted curve), $m=2$ (dashed curve), and $m=3$ (solid curve) for $a/M=0.8$. Bottom panel: The cases $a/M=0.866$ (solid curve), $a/M=0.917$ (dashed curve), and $a/M=0.954$ (solid curve) for $m=2$, in normalized units. Note, that here we show only the magnitude of $\psi_4$, factoring out its oscillations.}
\label{constancy}
\end{figure}

\setcounter{figure}{4}
\setcounter{table}{1}

This behavior affects Figures 5 and 6 in the original paper. The corrected figures are included. We also correct the values in Table II.

\begin{figure}[t]
\includegraphics[width=7.5cm]{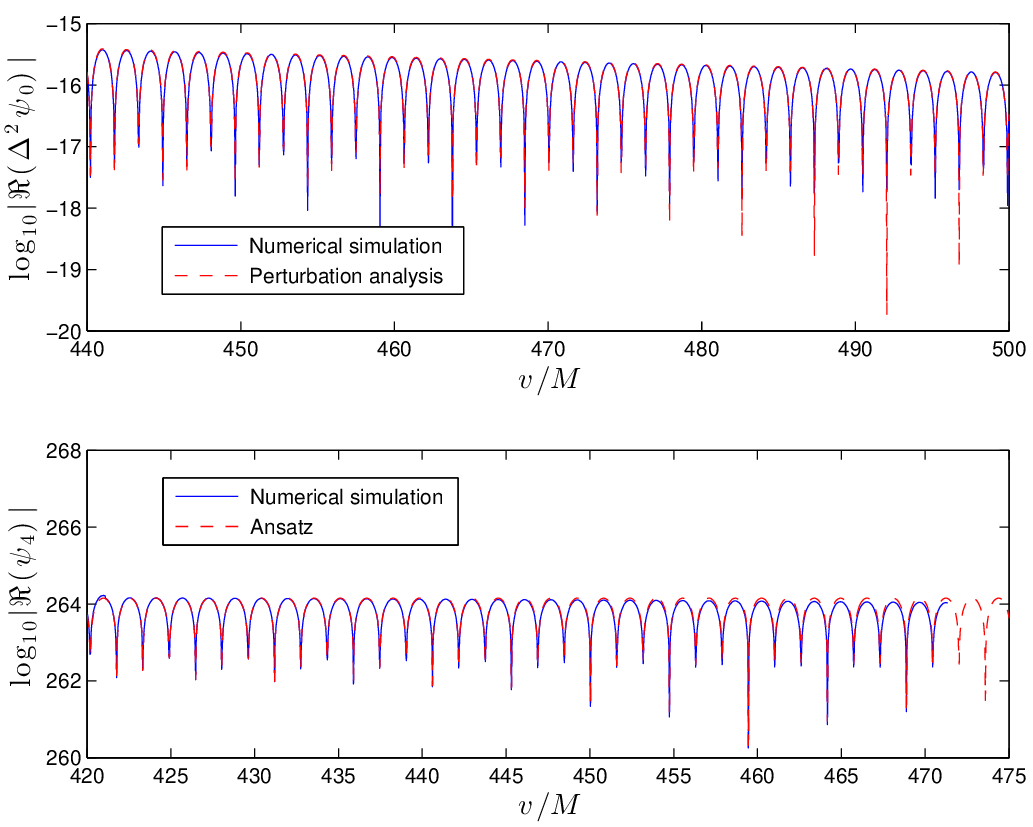}
\caption{The Weyl scalars $\Delta^2\psi_0$  in regularized coordinates (upper panel) and $\psi_4$ in ingoing Kerr coordinates (lower panel) as functions of $v/M$ along an outgoing null ray that intersects with the early part of the Cauchy horizon (as $v/M\to\infty$). For $\Delta^2\psi_0$ ($\psi_4$) we present in addition to our numerical results in a solid curve also the prediction of perturbation analysis (corrected Ansatz)  in a dashed curve.  
Here, $a/M=0.8$ and $m=2$ and the fields are on the equatorial plane.}
\label{psi04}
\end{figure}

\begin{figure}[h]
\includegraphics[width=7.5cm]{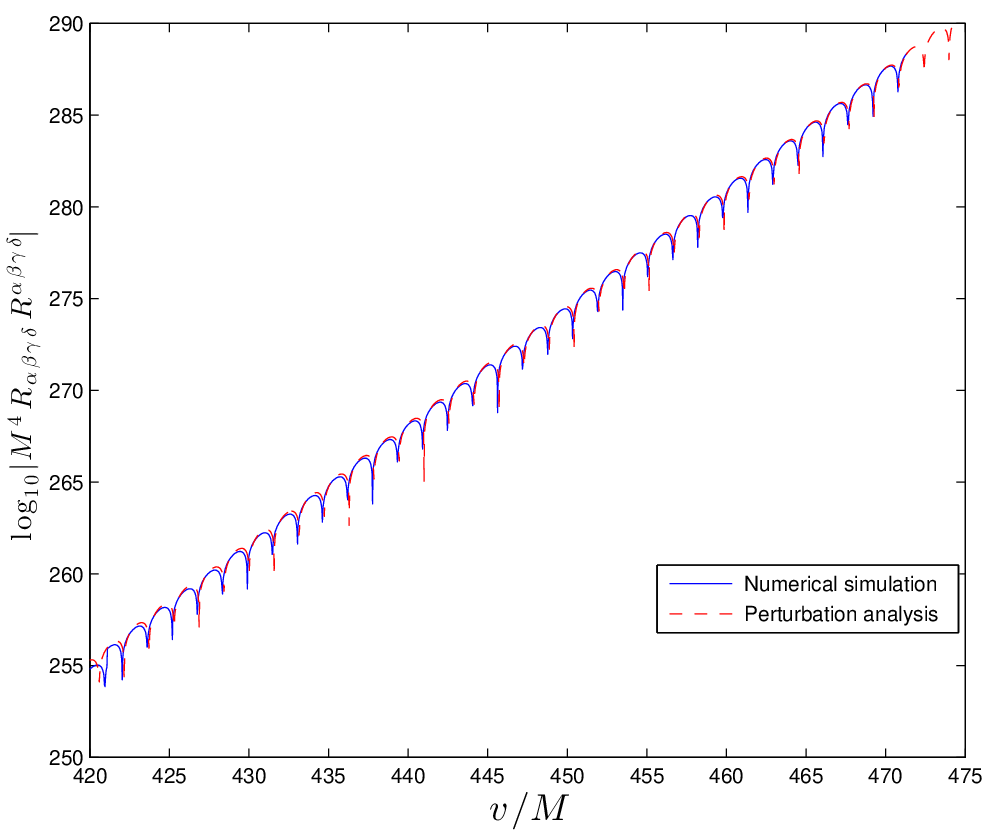}
\caption{The curvature scalar $R_{\alpha\beta\gamma\delta}R^{\alpha\beta\gamma\delta}$ as a function of $v/M$ along an outgoing null ray on the equatorial plane approaching the Cauchy horizon. Here, $a/M=0.8$ and $m=2$, and data are on the equatorial plane. The numerical simulation is shown with a solid curve, and the prediction of \cite{ori-99} with a dashed curve.}
\label{kretch}
\end{figure}

\begin{table}
\caption{Parameters for the numerically simulated $\psi_4$ approaching the Cauchy horizon: for $a/M=0.8$ for various values of $m$ (left side), and for $m=2$ for various values of $a/M$ (right side). The parameter $\omega$ is found from the numerical simulations. The relative difference between $\omega M$ and $m\Omega_-M$ is denoted by $\delta$. 
Data, on the equatorial plane, are presented in ingoing Kerr coordinates.}
\centering
\begin{tabular}{ || c | c | c | c || }
  \hline                       
  $m$    & $\omega M$ & $m\Omega_-M$ & $\delta$  \\ \hline 
  1 &   1.05 & 1.000  & 0.049 \\  \hline 
    2 &   2.09 & 2.000 & 0.044\\  \hline 
  3 &    3.14 & 3.000  & 0.046 \\
  \hline  
\end{tabular}
\begin{tabular}{ || c  | c | c  | c || }
  \hline                       
  $a/M$   & $\omega M$ & $m\Omega_-M$  & $\delta$ \\ \hline 
  0.800  & 2.09 &   2.000  & 0.044 \\  \hline 
    0.866  & 1.71 &  1.732 & 0.013 \\  \hline 
  0.917  & 1.48 &   1.528  & 0.032 \\ \hline  
   0.954 & 1.33 &   1.363  & 0.025 \\ \hline
    0.980  & 1.19 &   1.225 & 0.029 \\ \hline
\end{tabular}
\label{table:psi4}
\end{table}

\end{document}